\documentclass[11pt]{article}
\usepackage{graphicx}
\usepackage{epstopdf}%
\usepackage{amsmath,amssymb,dsfont,upgreek}

\usepackage[colorlinks=true,  linkcolor=blue, citecolor=blue, urlcolor=blue]{hyperref}

\newcommand{\ve}[1]{\ensuremath{\mathbf{#1}}}

\begin{document}

\title{Penultimate fate of a dirty-Fermi-liquid.}
\author{Alexander Punnoose and Alexander M. Finkel'stein}
%\date{October 14, 2009}
\maketitle

\newpage

\section{Dirty-Fermi-liquids}
%\subsection{General discussion}
It is generally expected that in the absence of strong external fields and when lattice effects can be ignored, weakly or even moderately interacting systems will remain a Fermi-liquid at sufficiently low temperatures. This simple expectation, we now understand, is wrong in two dimensions as even the weakest disorder has been shown to destabilize the Fermi-liquid at very low temperatures\,\cite{yellowbook,pedagogical}, driving the system into a strongly-correlated unusual metal or into an insulator\;\cite{punnoose05} depending on the initial (high temperature) resistance of the system. Indications are that the the two phases are kept apart by a quantum critical point (QCP) at a finite critical resistance\,\cite{punnoose05}.  We argue that the observed metal-insulator transition (MIT)\;\cite{kravoriginal} in two-dimensional electronic systems is a manifestation of this break down\,\cite{punnoose02,punnoose07}.

We are unfortunately very far from understanding the nature of the ultimate metallic state at absolute zero. We do, however, understand the process by which the Fermi-liquid is destabilized. In the diffusion regime, i.e., for temperatures (frequencies) less than the elastic scattering rate, $T\lesssim 1/\tau$, diffusion modes and not quasi-particles are the low lying propagating modes. Observable effects of quantum diffusion accumulate over long distances and long times and thereby develop only at much lower temperatures  $T\lesssim \frac{1}{\tau}e^{-\sigma}$, where $\sigma$ is the dimensionless conductance of the sample. Therefore, non-trivial effects are most clearly visible only in the vicinity of the transition where $\sigma\sim 1$.  To follow the effects over a wide range of temperature requires a reasonable value for $1/\tau$, putting practical constraints on the observability of quantum diffusion effects. The scattering rates are much too low in ultra-high mobility modulation doped or HIGFET structures ($1/\tau \ll 500$mK); silicon MOSFETs, on the other hand, have mobilities that are not too high leading to $1/\tau\sim 1$ - $2$\;K in the vicinity of the MIT, making them ideal playgrounds for studying quantum diffusion. It should be noted that in the absence of any other correlation-induced phase transition, all 2D systems will at sufficiently low temperatures be destabilized. 

The  theory of the dirty-Fermi-liquid which details the destruction of the Fermi-liquid phase up to and including the quantum critical point is reviewed here.

\subsection{Experimental systems}

The mobility of  2D quantum wells vary over orders of magnitude, ranging from $\sim 20,000$ - $40,000$ cm$^{2}/$Vs as in typical $n$-type Si-MOSFETs, to hundreds of thousands and even millions as in ultra-high mobility $p$-type GaAs heterostructures. The source and therefore the nature of the impurity potentials vary greatly between these systems.  It is, therefore, important to separate the effects of long versus short ranged random potential fluctuations on the scale of the interparticle distance, especially at very low carrier concentrations when the wavelength of the carriers is considerably increased. Screening of the impurity potentials becomes difficult in these ultra-high mobility systems; the resulting non-linear screening can induce non-trivial temperature dependencies in the transport properties of 2D systems and can in the extreme situation also lead to domain formations of high and low density regions\;\cite{fogler_theory} preventing us from experimentally studying the fundamental problem of the effects of random quantum scattering in a Fermi-liquid, whose full solution is unknown.

We argue below that MOSFETs provide an ideal environment to study the properties of a 2D dirty-Fermi-liquid. The moderate values of the mobility of these systems  results in an impurity elastic scattering rate slightly larger than a Kelvin. Comparing with the Drude conductance $\sigma = n\mu$, expressed in units of $e^2/h$, where $n$ is the electron density and $\mu$ the mobility, moderately high mobility implies that the density in the vicinity of the quantum conductance is of the order of $10^{11}$\;cm$^{-2}$. The corresponding Fermi-temperature is of the order of $10$\;K, and is therefore degenerate already at not so low temperatures. Also, because the  electron-phonon coupling in silicon is weak due to the presence of inversion symmetry in the bulk, electron-phonon scattering contribution to the resistivity is negligible at low temperatures. Finally, another important characteristic of MOSFETs, in strong contrast to heterostructures, is that the elastic single particle scattering rate  is comparable to the transport scattering rate; they typically differ by a factor of $10$ in other systems. Quantum scattering therefore is the dominant scattering mechanism in MOSFETs.    

The Coulomb interaction energy at the densities of interest are typically $10$ times larger than the Fermi-energy, i.e.,  $r_s=E_c/E_F\approx 10$, where $E_c$ is the Coulomb energy of neighboring electrons. At these $r_s$ values, detailed  numerics indicate no phase transitions, at least up to $r_s\approx 25$ where a paramagent to ferromagnet transition is found to occur in a system with single valley (no transition has been found in a two-valley system)\;\cite{senatore05,senatore09}. At still higher $r_s\approx 35$, the electron liquid is expected to form a Wigner crystal, although none has been observed experimentally\;\cite{tanatar89}. These studies suggest a Fermi-liquid with moderately strongly interacting quasi-particle scattering amplitudes around $r_s\approx 10$.

MOSFETs have the additional advantage that they are a multi-valley system. The conduction band of an $n$-(001) silicon inversion layer has two almost degenerate valleys located close to the $X$-points in the Brillouin zone\;\cite{ando}. The presence of two equivalent valleys further enhances the correlation effects\;\cite{punnoose02,punnoose07} making MOSFETs ideal systems to study the effects of quantum disorder scattering in a Fermi-liquid. In the following we consider the number of equivalent valleys to be equal to $N$. 

\section{Diffusion regime} 
%\subsection{General discussion}
Fermi-liquid renormalizations, like mass, g-factor, or equivalently, specific heat and spin-susceptibility, occur on the scale $\lambda_F$. As a result, the quasi-particle description remains valid when  the elastic mean free path, $l \gg \lambda_F$. This translates to $E_F\gg \hbar/\tau$, where $\tau$ is the elastic scattering time related to $l$ by the Fermi-velocity.  Multiple impurity scattering leads to the diffusive propagation of the quasi-particles for times $\gg \tau$ or frequencies $\omega \ll 1/\tau$, characterized by the quasi-particle diffusion constant, $D$. Fermi-liquid interactions between the quasi-particles, however, modifies $D$ in a scale  (frequency/temperature) dependent way, making the conductivity $\sigma$, which is related to $D$ through Einstein's relation  $\sigma=(\partial n/\partial \mu)D$, temperature dependent.   Altshuler, Aronov and Lee showed\;\cite{AAbook,AAL} that the corrections are localizing in the limit of weak quasi-particle interactions, so that the electronic system becomes an Anderson insulator as the temperature $T\rightarrow 0$. Here, $\partial n/\partial \mu$ is the thermodynamic density of states which is not renormalized in the diffusion limit. (Experimentally, the compressibility is found to change sign in the strong-localization regime\;\cite{jiang00,yacoby01}, which is outside the scope of the diffusion theory.)

In the discussion above  the Fermi-liquid scattering amplitudes were considered to be scale independent, as is the case in a clean Fermi-liquid. It was shown, however, that the disorder averaged amplitudes are in fact scale dependent at low energies, $\omega\lesssim 1/\tau$,  due to the diffusive propagation of the quasi-particles\;\cite{sasha83}.  For weak disorder (low resistance) the effective scattering amplitudes were shown to become strongly correlated over long distances and eventually diverging  at a very small but finite scale (temperature) $T^*$.  The enhanced correlation overcome localization so that the resistance $\rho^*$ at $T^*$ is finite. This interplay of diffusion and interaction controls the  physics of dirty-Fermi-liquid down to the scale $T^*$, below which the liquid is no longer describable by diffusing Fermi-liquid quasi-particles. 

In contrast to the weak disorder case, it was shown recently\;\cite{punnoose05} that the situation is very different  when the disorder (resistance) is increased. For not so weak disorder the Fermi-liquid correlations were shown to be suppressed at long distance, so that  the dirty-Fermi-liquid becomes an Anderson-like insulator at low temperatures. It was also established in Ref.\;\cite{punnoose05},  in the framework of an effective large-$N$ dirty-Fermi-liquid theory, that a QCP at a finite critical resistance separates the metallic and the insulating phases, implying the existence of a MIT in two dimensions.  The theoretical details of these results  and their experimental tests are briefly discussed below.

\subsection{Theory: Background}

The minimal model that describes the problem of quasi-particle diffusion in the
field of impurities and interactions is the
non-linear sigma model\;\cite{sasha83}

\begin{subequations}
\begin{eqnarray}
S[Q]&=&S_\textrm{dis}+S_\textrm{int}\\
S_\textrm{dis}&=&-\frac{\pi \nu}{8}\int d^2\ve{r}\ \textrm
{Tr}\left[ D(\nabla Q)^2 -4z (\hat{\epsilon}Q)\right]\label{eqn:action_dis}\\
S_\textrm{int} &=&+\frac{\pi \nu}{8}\int d^2\ve{r}\
Q(\hat{\Gamma}_1+\hat{\Gamma}_2+\hat{\Gamma}_c) Q\label{eqn:action_ee}
\end{eqnarray} \label{eqn:action}
\end{subequations}

The functional $S[Q]$ describes interacting quasi-particles with
energies $\epsilon<1/\tau$. ($\hat{\epsilon}$ is the energy matrix whose elements involve the energy of the quasi-particles.)  Note that the region of higher energies
$1/\tau < |\epsilon | \lesssim E_F$, containing Fermi-liquid
renormalizations of the clean liquid are incorporated as the input
parameters into this theory.
The $Q$-field here is an auxiliary matrix field describing the
impurity scattering. It satisfies the constraints: $Q^2 = 1,
Q=Q^\dagger,$ and Tr\;$Q =0$. The fluctuations of $Q$ in the
particle-hole (diffusons) and particle-particle (cooperons) channels
are described by the diffusive propagators
$\mathcal{D}(q,\omega_n)\sim 1/(Dq^2+z\omega_n)$ (we use Matsubara
frequencies). These propagators capture the diffusive evolution  of
the charge and spin density fluctuations at large times and length
scales, i.e., small frequency and wave-vectors. The parameter $z$ is the frequency renormalization parameter\;\cite{sasha83,sasha84}, which can be interpreted as the renormalization of the density of states of the diffusion modes ($z=1$ for non-interacting electrons), and as a result controls the thermodynamic properties of the dirty-Fermi-liquid. The cooperons
capture the effects of quantum interference, which at the
perturbative level leads to the well known weak-localization
corrections.

The  quasi-particle interactions are characterized by the Fermi-liquid amplitude matrices, $\hat{\Gamma}_1$ and $\hat{\Gamma}_2$,  in the diffuson channel and, $\hat{\Gamma}_c$, in the cooperon channel, respectively. The details of their matrix structure in the space of spin,  frequency and momentum are not shown here, they can be found in Ref.\;\cite{yellowbook}. (See Ref.\;\cite{punnoose09a} for their valley dependence.)
The magnitudes of the amplitudes, ${\Gamma}_1$ and ${\Gamma}_2$,  are related to the standard zeroth order harmonics of the Fermi-liquid amplitudes
$F_0^{s,a}$  as $\Gamma_{1,2}=-F_0^{s,a}/(1+F_0^{s,a})$. 

The scaling of the parameters, $D, z$ and $\Gamma_i$'s ($i=1,2$), together describe the transport and thermodynamic properties of the dirty-Fermi-liquid.  Special care is, however, required when  Coulomb interactions are present\;\cite{sasha83}. The $\Gamma_{1}$ amplitude in this case includes amplitudes of the kind  which can be separated by cutting the statically screened long-ranged Coulomb line once. They are usually denoted as $\Gamma_{0}$.  This distinction is important because the polarization operator which is irreducible to cutting a Coulomb line does not include $\Gamma_{0}$. Conservation laws then require that $z=2N(\Gamma_0+\Gamma_1)-\Gamma_2$, where $N$ is the number of valley degrees of freedom\;\cite{sasha83,punnoose09a}. 

As discussed in the last section, the diffusion constant and the Fermi-liquid scattering amplitudes become scale dependent in a dirty-Fermi-liquid. It has been shown that renormalization group (RG) theory applied to the dirty-Fermi-liquid model given in (\ref{eqn:action}) is able to capture the scale dependence originating from the interplay of disorder and interactions to all orders in the interaction amplitude, making it the most promising analytical tool available to understand the physics of disordered systems.  Pedagogical reviews  of the RG theory can be found in Refs.\;\cite{yellowbook,pedagogical}. For more recent advances, see Ref.\;\cite{pruiskin}.  

The scaling variables are described in terms of the dimensionless parameters, the resistance $\rho=1/2N(2\pi^2\nu D)$, and the interaction  $\Gamma_2$ and $z$. It follows, however, from the structure of the action in (\ref{eqn:action})  that the variables can be expressed as $\rho$ and $\gamma_2=\Gamma_2/z$ which does not contain z explicitly. In terms of these variables, the equations for  $\rho$ and $\gamma_2$ form a closed set of  equations  independent of $z$. In general, they take the form:

\begin{subequations}
\begin{eqnarray}
\frac{d\ln\rho}{d\xi}&=&\rho \beta_\rho(\rho,\gamma_2)\\
\frac{d\gamma_2}{d\xi}&=&\rho \beta_{\gamma_2}(\rho,\gamma_2)\\
\frac{d\ln z}{d\xi}&=&\rho\beta_z(\rho,\gamma_2)
\end{eqnarray}
\end{subequations}
The scale $\xi=\log(1/T\tau)$, where to logarithmic accuracy $1/\tau$ is used as the upper cut-off. Note that the diffusive contribution involving the parameter $\gamma_1$ is universal when the dynamically screened Coulomb interaction is involved, i.e., when  $\gamma_0=\Gamma_0/z$ is included. Hence it does not appear as an explicit parameter in these equations\;\cite{AAbook}. It is a well established result with great importance for the general structure of the theory that the static amplitude $\Gamma_0$ does not acquire diffusion corrections and is therefore not renormalized. When combined with the conservation laws discussed above it follows  that $d(2N\Gamma_1-\Gamma_2)=dz$, from which one may extract the renormalization of $\Gamma_1$\;\cite{sasha83}. The above discussions imply that the electronic transport ($\rho$) and thermodynamic (involving $z$) properties are describable by a two-parameter scaling theory involving $\rho$ and $\gamma_2$ only. (For repulsive interactions, the amplitude $\gamma_c$ scales to a resistance dependent value for finite $N$ and hence is not considered here. The situation in the $N\rightarrow\infty$ limit is different and will be discussed later.)

The $\beta$-functions are multiplied by a factor $\rho$ to emphasize that the sought after corrections appear as a result of the underlying diffusive propagators. The complete form of the $\beta$-functions are unknown. The general approach, however, is to  expand the functions in a power series in $\rho$ as $\beta(\rho,\gamma_2)\sim  \beta_1(\gamma_2)+\rho\beta_2(\gamma_2)+\cdots$, such that for each power of $\rho$ the full dependence on $\gamma_2$ is obtained. This is possible, in principle, because the maximum number of allowed interaction vertices are limited (before the ladder sums are included) by the number of momentum integrations involving the diffusive propagators, and since each propagator gives a factor of $1/D\sim \rho$, for a given order in $\rho$ only a finite number of vertices contribute.  So far, only the first order (one-loop) results have been obtained exactly\;\cite{sasha83}. Such a solution has the obvious limitation that it is applicable only when $\rho \ll 1$ and is not applicable near $\rho\sim 1$ where the MIT is experimentally found to be located\;\cite{kravoriginal}. Therefore to approach the MIT, the disorder has to be treated beyond one-loop (at least to two-loop order) while adequately retaining the effects of interaction. Exact results to order $\rho^2$ have not been obtained so far. It was shown, however, that certain simplifications that come with standard large-$N$ theories can be exploited to obtain the two-loop results\;\cite{punnoose05}. The RG equations to two-loop order within the large-$N$ approximation showed, for the first time, the existence of a QCP that describes the MIT separating the metallic from the insulating phase in 2D.  Analysis of the QCP revealed a non-Fermi-liquid fixed point with diverging specific heat and spin-susceptibility. The one and two-loop RG equations  are presented below and comparison with experiments are discussed.

\subsection{Finite-$N$: One-loop}
%\subsubsection{RG equations}
The theoretical details of the evolution of $\rho$ and $\gamma_2$ in this 
region with $\rho\ll 1$ were discussed in detail in Ref.\;\cite{punnoose02}. For completeness we give the equations and only discuss its salient features here. The RG equations generalized to include $N$ equivalent valley degrees of freedom are:
\begin{subequations}\label{eqn:oneloop}
\begin{eqnarray}
\frac{d\ln\rho}{d\xi}&=&\rho\;\left[N+1-(4N^2-1)\Phi(\gamma_2)
\right]
\label{eqn:onelooprho}\\
\frac{d\gamma_2}{d\xi}&=&\frac{\rho}{2}\;(1+\gamma_2)^2
\label{eqn:oneloopgamma2}\\
\frac{d\ln z}{d\xi}&=&
\frac{\rho}{2}\left[-1+(4N^2-1)\gamma_2\right]\label{eqn:oneloopZ}
\end{eqnarray}
\end{subequations}
where $\Phi(\gamma_2)=(1+1/\gamma_2)\ln(1+\gamma_2)-1$. The factor $(4N^2-1)$ corresponds to the number of spin-valley triplet channels; the factor $N$ in Eq.\;(\ref{eqn:onelooprho}) corresponds to the weak-localization corrections, which is enhanced due to the $N$ extra degrees of freedom. The  factor of one in (\ref{eqn:onelooprho}) is the contribution of the long-ranged Coulomb singlet-amplitude, which after dynamical screening is universal; it should be emphasized that the factor of one arising in equation (\ref{eqn:oneloopgamma2}) for $\gamma_2$ has the same origin, and as a result, setting the initial value of $\gamma_2$ to zero does not switch the interactions off.   

The following salient features are to be noted: (1) while the amplitude $\gamma_2$ increases monotonically as the temperature is reduced; (2) the resistance, as a result, has a characteristic non-monotonic form changing 
from insulating behavior ($d\rho/dT<0$) at high temperatures to metallic behavior ($d\rho/dT>0$) at low temperatures. The change in slope occurs at a maximum value $\rho_\textrm{max}$ at a temperature  $T =T_\textrm{max}$, neither of which are universal; (3) the corresponding value of the amplitude $\gamma_2$ at $T_\textrm{max}$ is, however, universal at the one-loop order, depending only on $N$; for $N =1$, it is $2.08$, 
whereas for $N =2$, it has the considerably lower value $0.45$; (4) it follows from the form of the equations that the temperature dependence of $\rho(T)/\rho_\textrm{max}$ and $\gamma_2(T)$ are universal when  plotted as functions of $\rho_\textrm{max}\ln(T/T_\textrm{max})$. 
%\subsubsection{Experiment verification of universal scaling}

As explained in the introduction, the theory described here is applicable only in the diffusive regime. It is critical that the diffusive regime (temperature and density range) is identified carefully before the scaling is tested. In Ref.\;\cite{punnoose07} a method that takes advantage of the $B$ and $T$ dependence of the magnetoconductance, $\sigma(B,T)$,  in parallel field was exploited to identify the diffusive regime. It is known that while $\sigma(B,T)$ scales as $(B/T)^2$ in the diffusive regime\;\cite{tvr,castellani98}, in the  ballistic regime it scales as $B^2/T$\;\cite{zala_2002}. This distinction was used to reliably identify the diffusive regime in a MOSFET sample. The experimental details and the results of the comparison with theory can be found in Ref.\;\cite{punnoose07}, where for the first time the scaling of the interaction amplitude was established. Remarkably,  the parameter $\gamma_2$ at $T=T_\textrm{max}$ was found to correspond to  0.45 as predicted by theory for $N=2$.  

\subsection{Two-loops and the MIT in the limit $N\rightarrow\infty$ }

It is instructive to analyze Eq.~(\ref{eqn:oneloop}) for $N\gg 1$.
In this case it is easily seen that the value of the interaction amplitude $\gamma_2$ at $T=T_\textrm{max}$, i.e., when $d\rho/dT=0$ in Eq.~(\ref{eqn:onelooprho}), scales as 
$\gamma_2(T_\text{max};N)\approx 1/2N$ for large $N$. Furthermore, because the
presence of $N$-valleys increases the number of conducting channels,
the resistance scales as $\rho\sim 1/N$. This can be explicitly
taken into account by defining the maximum resistance, which is the
only free parameter at the one-loop level, as $\rho_\text{max}=1/N$.
Defining new variables $t=\rho(T)/\rho_\text{max}$ and
$\theta_2=\gamma_2(T)/\gamma_2(T_\text{max};N)$, and keeping in mind
that $N\gg 1$, Eq.~(\ref{eqn:oneloop}) can be expanded in powers of
$1/N$:
\begin{subequations}\label{eqn:largeNoneloopexpansion}
\begin{eqnarray}
\frac{d\ln t}{d\eta}&=&t \left[1-\theta_2 +
\mathcal{O}(\theta_2/N)\right]
\label{eqn:largeNonelooprhoexpansion}\\
\frac{d\theta_2}{d\eta}&=&t\left[1+\mathcal{O}(\theta_2/N)\right]
\label{eqn:largeNoneloopgamma2expansion}\\
\frac{d\ln z}{d\eta}&=&
t\left[\theta_2-\mathcal{O}(1/N)\right]\label{eqn:largeNoneloopZexpansion}
\end{eqnarray}
\end{subequations}
After rescaling the temperature scale becomes $\eta=\ln(T_\text{max}/T)$. It
is now possible to take the limit $N\rightarrow \infty$, keeping $t$
and $\theta_2$ finite. Note that the parameter $t$ corresponds to
the resistance per valley, $t=1/(2\pi)^2\nu D$, with $\nu$ being the
density of states of a single spin and valley species. In the
infinite-$N$ limit Eq.~(\ref{eqn:largeNoneloopexpansion}) reduces to
\begin{subequations}\label{eqn:largeNoneloop}
\begin{eqnarray}
\frac{d\ln t}{d\eta}&=&t (1-\theta_2)
\label{eqn:largeNonelooprho}\\
\frac{d\theta_2}{d\eta}&=&t
\label{eqn:largeNoneloopgamma2}\\
\frac{d\ln z}{d\eta}&=& t\theta_2\label{eqn:largeNoneloopZ}
\end{eqnarray}
\end{subequations}
The form of Eq.~(\ref{eqn:largeNoneloop}) illustrates the fact that in the
$N\rightarrow\infty$ limit the maximum number of interaction amplitudes that
can appear are limited by the power of $t$. This is to be compared
with the function $\Phi(\gamma_2)$ appearing in
Eq.~(\ref{eqn:oneloop}). Clearly, this is one of the main
simplifications provided on taking the $N\rightarrow\infty$.

Taking advantage of this simplification, the one-loop results above were extended to two-loops, i.e., order $t^2$ in Ref.\;\cite{punnoose05}. It is easy to see that, unlike Eqs.\;(\ref{eqn:oneloop}), the solutions of Eqs.\;(\ref{eqn:largeNoneloop}) are valid up to $T=0$, at which point $\theta_2\rightarrow\infty$ and $t\rightarrow 0$, providing an asymptotically exact theory all the way down to $T=0$. (As discussed earlier, $\gamma_2$ for finite $N$ diverges at a certain finite temperature $T^*$ below which the RG equations are no longer valid.) It is therefore left to analyze the large $t$ behavior at two-loop order.

The RG equations in the $N\rightarrow\infty$ limit are given as\;\cite{punnoose05}:
\begin{subequations}\label{eqn:twoloop}
\begin{eqnarray}
\frac{d\ln t}{d\eta}&=&t(\alpha-\Theta)+t^2(1-\alpha-\alpha\Theta+c_t\Theta^2)\label{eqn:twoloop-t}\\
\frac{d\Theta}{d\eta}&=&t(1+\alpha+\alpha\Theta)-4t^2\left[(1+\alpha)\Theta+\frac{\alpha}{2}\Theta^2+c_\Theta\Theta^3\right]\label{eqn:twoloop-theta}\\
\frac{d\ln z}{d\eta}&=&t\Theta\label{eqn:twoloop-z}
\end{eqnarray}
\end{subequations}
where $c_t=(5-\pi^2/3)/2$ and $c_\Theta=(1-\pi^2/12)/2$. The parameter $\alpha$ takes values $0$ and $1$, depending on if the cooperon channel is absent or present, respectively. In the  large-$N$ limit because rescattering in the cooper channel is not accompanied by factors of $N$ such processes are irrelevant and hence $\theta_c=N\gamma_c$  appears on equal footing as $\theta_2$. (Violation of time-reversal by a magnetic field for example will, however, suppress the contribution of $\gamma_c$.) The amplitudes $\theta_2$ and $\theta_c$ always appear in the combination  $\Theta=\theta_2+\alpha\theta_c$. 

It is clear from Eq.\;(\ref{eqn:twoloop-theta}) that for large enough resistance $d\Theta/d\eta <0$. Suppression of the $\Theta$ amplitude leads to allows $t$ to grow to large values (insulator) as the temperature is lowered. This is a clear departure from the metallic behavior observed for small $t$. The insulating and metallic phases are then separated by a fixed point given at $(t_c, \Theta_c)$ where $d\ln t/d\eta=0$ and $d\Theta/d\eta=0$. The quantum critical nature of the fixed point is apparent from the equation for $z$ which at $(t_c,\Theta_c)$ diverges as $z\sim 1/T^{t_c\Theta_c}$.  The parameter  $z$, which corresponds to the frequency renormalization, can be interpreted as the density of states of the underlying diffusion modes. Therefore it controls the behavior of the specific heat $C_v\sim (z\nu)T$. Hence, divergence in $z$ implies a divergence in the specific heat constant $C_v/T$ signaling the break down of the Fermi-liquid behavior at the critical point. Similar divergence is expected in the Pauli spin susceptibility also. Since the interaction parameter $\Theta$ is finite at the critical point, the divergence in the Pauli spin susceptibility is not related to any Stoner-like magnetic instability.

The feature that can be compared with experiments comes by observing in Eqs.\;(\ref{eqn:twoloop}) that unlike in the metallic phase where $dt/d\eta < 0$ and $d\Theta/d\eta >0$, in the insulating phase the opposite behavior is obtained, namely, $dt/d\eta >0$ and $d\Theta/d\eta <0$. In Ref.\;\cite{punnoose07}, precisely such a scaling was confirmed experimentally.

\section{Conclusions}
We believe thereby that the two-parameter theory of the dirty-Fermi-liquid reviewed here captures  both quantitatively (for small disorder) and qualitatively (for large disorder) the physics of the MIT in 2D. 

\section{Acknowledgments}
This work was partially supported by the US-Israel Binational Science Foundation grant 2006375. The work at City College was supported by DOE grant DOE-FG02-84-ER45153.

%\subsection{Comparison with experiments}

\bibliographystyle{unsrt}

%\bibliography{MITbib}

\end{document}